\documentclass[12pt]{iopart}

\usepackage{iopams}  

\usepackage{amssymb}
\usepackage[dvips]{color}
\usepackage{epsfig}
\usepackage{dcolumn}
\usepackage{bm}

\def\a{\alpha}

\def\g{\gamma}
\def\G{\Gamma}
\def\d{\delta}
\def\D{\Delta}
\def\e{\epsilon}

\def\s{\sigma}

\def\wid{\widetilde}

\begin{document}

\title{Spin polarized transport driven by square voltage pulses in a quantum dot system}

\author{F. M. Souza and J. A. Gomez}

\address{International Centre for Condensed Matter
Physics, Universidade de Bras{\'i}lia, 70904-910, Bras{\'i}lia-DF,
Brazil}
\ead{fmsouza@unb.br}

\begin{abstract}
We calculate current, spin current and tunnel magnetoresistance (TMR) for a quantum dot coupled to ferromagnetic leads in the presence of a square wave of bias voltage. Our results are obtained via time-dependent nonequilibrium Green function. Both parallel and antiparallel lead magnetization alignments are considered. The main findings include a wave of spin accumulation and spin current that can change sign as the time evolves, spikes in the TMR signal and a TMR sign change due to an ultrafast switch from forward to reverse current in the emitter lead. 
\end{abstract}

\maketitle

\section{Introduction}

Spintronic\cite{spintronics} has proved to be of great technological importance with the development of memory storage devices and magnetic sensors based on giant magnetoresistance (GMR)\cite{gmr} and tunnel magnetoresistance (TMR).\cite{tmr} More recently, one of the major achievements in the spintronic field was the coherent control of single electron spins in quantum dots.\cite{th03,jme04,jrp05,mvgd05,fhlk06,ag06,mvgd06,mhm07,sn07,mk07} Such a control is a fundamental step toward the generation of quantum bits based on the electron spin for further implementation of quantum computers.\cite{dl98,ai99,gb99} There are a few recently developed techniques to manipulate coherently the electron spin in quantum dots. In those techniques square pulses of bias or gate voltages are applied in order to prepare and measure spin states.\cite{seejme04} This turns quite desirable the study of quantum transport in the presence of pulses of bias voltage. For single level quantum dots coupled to nonmagnetic leads, it is well known that coherent oscillations (ringing) of the current appear when a step like bias voltage is applied across the junction.\cite{apj94} In the presence of ferromagnetic leads and/or Zeeman-split level, though, it was found that the ringing response of the current develops spin fingerprints.\cite{fms07_3,ep08} Additionally, polarized current spikes can be generated when a bias voltage is abruptly turned off.\cite{fms07_2}

\begin{figure}[b]
\par
\begin{center}
\epsfig{file=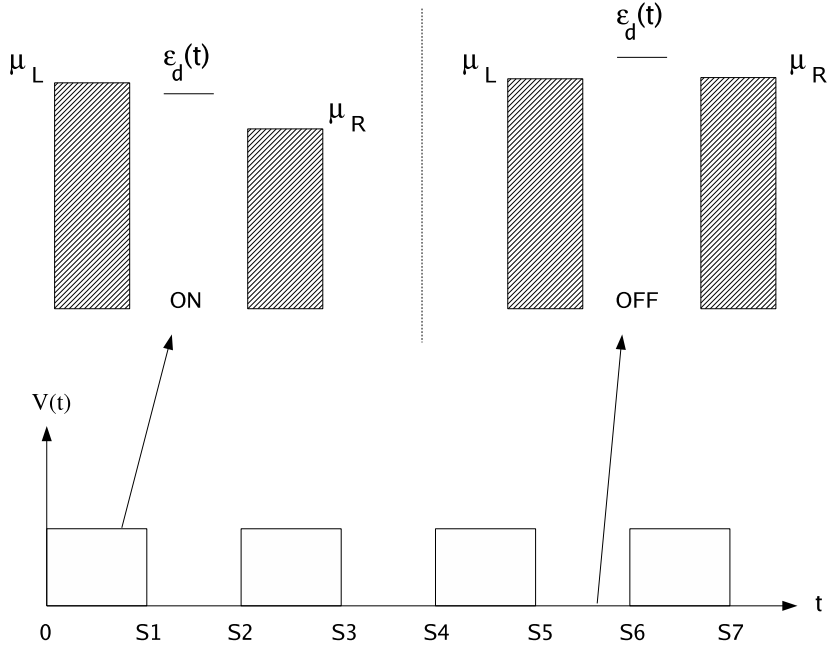, width=0.8\textwidth}
\end{center}
\caption{Schematic energy diagram for the system considered. The dot level $\e_d(t)$ is coupled to a left (emitter) and to a right (collector) lead with chemical potentials $\mu_L$ and $\mu_R$, respectively. During a pulse of bias voltage the level $\e_d(t)$ and the right chemical potential $\mu_R$ are shifted down, thus allowing charge and spin transport through the system. When the bias is turned off the level $\e_d(t)$ becomes above both $\mu_L$ and $\mu_R$, which results in a transient discharging of the dot. This discharge process is spin-dependent due to the ferromagnetic leads.} \label{fig1}
\end{figure}

Systems with high TMR values are of current interest in spintronics, motivated mainly by magnetic memory and sensor applications.\cite{tmrsensors} Recently, it was found relatively high TMR values (80\%-200\%) in magnetic tunnel junction (MTJ) consisting of Fe and Co spaced by an oxide layer.\cite{sspp04,sy04,hightmr} In particular, if instead of an insulator layer we have a quantum dot sandwiched by two ferromagnetic leads,\cite{expFMdotFM} it is possible to have additional effects, like spin-accumulation and strong Coulomb interaction that result in a more wealth physics of the TMR response.\cite{iw07,kw06,iw06_2,iw05_2,fms04,rl03,wr03,wr01,jk03,mb04}

In the present work we extend the recent study in Ref. \cite{fms07_2} by considering a sequence of square pulses of bias voltage instead of only a single pulse. A study
on electron spin dynamics when a sequence of voltage pulses is applied in a quantum dot system, was recently developed by Stefanucci.\cite{gs08} Since square waves are of fundamental importance for the conventional electronics (e.g. digital switching circuits, synchronous logic circuits, binary logic devices)\cite{booksdigital} we believe that it is relevant to consider the interplay between square waves of bias voltage and spin dependent phenomena. The calculation presented here is based on the nonequilibrium Green function technique. The main finding is a wave-like behavior of the spin accumulation, spin current and TMR, that can switch sign periodically as the time evolves. In particular, the TMR develops a periodic singularity that results in relatively high TMR and a fast switching of its sign.

The paper is organized as follows. In Sec. II we derive an expression for the current, in Sec. III we show and discuss the results and in Sec. IV we conclude.

\section{Hamiltonian and Transport Formulation}

To describe the system illustrated in Fig. \ref{fig1} we use the following Hamiltonian
\begin{eqnarray}\label{hamil}
H&=&\sum_{\mathbf{k} \s \eta} \e_{\mathbf{k} \s \eta}(t)
c_{\mathbf{k} \s \eta}^\dagger c_{\mathbf{k} \s \eta}+\sum_{\s}
\e_d(t) d_\s^\dagger d_\s\nonumber\\
&\quad&+\sum_{\mathbf{k} \s \eta} (V
c_{\mathbf{k}\s \eta}^\dagger d_\s +V^*d_\s^\dagger c_{\mathbf{k} \s \eta}),
\end{eqnarray}
where $c_{\mathbf{k} \s \eta}$ ($c_{\mathbf{k} \s \eta}^\dagger$)  and $d_\s$ ($d_\s^\dagger$) are the annihilation (creation)
operators for electrons in the lead $\eta$ and in the dot, respectively. The energies $\e_{\mathbf{k} \s \eta}(t)$ and $\e_d (t)$ are the time-dependent
energies for the electrons in lead $\eta$ ($\eta=$L or R for left or right) and in the dot, respectively. The labels
$\mathbf{k}$ and $\s$ denote the electron wave vector and the spin, respectively. More explicitly, these energies are written as
$\e_{\mathbf{k} \s \eta}(t)=\e_{\mathbf{k} \s \eta}^0+ \Delta_\eta
(t)$ and $\e_d (t)=\e_d^0 + \Delta_d (t)$, where $\e_{\mathbf{k} \s \eta}^0$ and $\e_d^0$  are time-independent energies and $\Delta_{\eta,d} (t)$ gives the time evolution of the external bias. Using Eq. (\ref{hamil}) inside the current definition $I_\s^\eta(t)=-ie \langle [H,N_\s^\eta] \rangle$, where $e$ is the electron charge ($e > 0$) and $N_\s^\eta = \sum_{\mathbf{k}} c_{\mathbf{k} \s \eta}^\dagger c_{\mathbf{k} \s \eta}$, we can show in the noninteracting\cite{commentUzero} case and wideband limit that\cite{hh96,gs04}
\begin{eqnarray}\label{Iseta}
I_\s^\eta(t)=- e \G_\s^\eta \{ \langle n_\s (t) \rangle + \int
\frac{d\e}{\pi} f_\eta (\e) \mathrm{Im} [A_{\s \eta}(\e,t)]\},
\end{eqnarray}
where $f_\eta(\e)$ is the Fermi distribution function for lead $\eta$, and $\langle n_\s (t) \rangle$ is the time-dependent dot's
occupation, given by
\begin{eqnarray}\label{nst}
\langle n_\s(t) \rangle &=& \mathrm{Im} \{ G_{\s\s}^< (t,t)
\}\nonumber\\&=&\sum_\eta \G_\s^\eta \int \frac{d\e}{2\pi} f_\eta (\e) |A_{\s
\eta} (\e,t)|^2.
\end{eqnarray}
The function $A_{\s \eta}(\e,t)$ is defined as
\begin{equation}\label{Averyfirst}
A_{\s \eta}(\e,t)=\int_{-\infty}^{t} dt_1 G_{\s\s}^r(t,t_1) e^{[i
\e (t-t_1) - i \int_{t}^{t_1} d\wid{t} \D_\eta (\wid{t})]},
\end{equation}
where the retarded Green function in the noninteracting model is given by
\begin{equation}\label{Gr}
G_{\s\s}^r(t,t_1)=-i \theta(t-t_1) e^{-\frac{\G_\s}{2}(t-t_1)}
e^{-i \int_{t_1}^{t} d\wid{t} \e_d(\wid{t})},
\end{equation}
with $\G_\s=\G_\s^L+\G_\s^R$ and $\e_d=\e_0+\D_d(t)$. The quantities $\G_\s^L$ and $\G_\s^R$
give the tunneling rate between left and the right leads into/out the dot, respectively.
Substituting Eq. (\ref{Gr}) into Eq. (\ref{Averyfirst}) we find
\begin{eqnarray}\label{Ageral}
A_{\s \eta}(\e,t)&=&-i \int_{-\infty}^{t} dt_1 e^{i( \e
+i\frac{\G_\s}{2})(t-t_1)} e^{- i \int_{t_1}^{t} d\tilde{t}
[\e_d(\tilde{t})-\D_\eta (\tilde{t})]},\nonumber\\&=& -i e^{i( \e
+i\frac{\G_\s}{2})t} \times \nonumber \\ && \phantom{xx} \int_{-\infty}^{t} dt_1 e^{-i\{( \e
+i\frac{\G_\s}{2})t_1 + \int_{t_1}^{t} d\tilde{t}
[\e_d(\tilde{t})-\D_\eta (\tilde{t})]\}}.
\end{eqnarray}
Solving Eq. (\ref{Ageral}) for a bias voltage of the kind
\begin{equation}
 V_{\eta/d}(t)=V^0_{\eta/d} \phi(t) =V^0_{\eta/d} \sum_{n=1}^{\infty} \theta(t-s_{n-1}) \theta(s_n-t),
\end{equation}
with $\D_\eta(t)=-V_\eta$ and $\D_d(t)=-V_d(t)$, we obtain
\begin{eqnarray}\label{Awithint}
 &&A_{\s \eta}(\e,t)=-i \int_{-\infty}^{s_0} dt_1 \g(t,t_1) e^{-i \a_\eta [t \phi(t) + \sum_{n=0}^{N} (-1)^{n+1}s_n]}\nonumber\\&&
-i\xi \sum_{n=0}^{N-1} \int_{s_n}^{s_{n+1}} dt_1 \g(t,t_1) e^{-i \a_\eta [t \phi(t) + \sum_{l=n+1}^{N} s_l -t_1 \phi(t_1)]}\nonumber \\ &&
-i \int_{s_{N}}^t dt_1 \g(t,t_1) e^{-i \a_\eta (t-t_1) \phi(t)},
\end{eqnarray}
where $N$ gives the last instant $t_N$ (time in which $V(t)$ is turned on or off) before the time $t$ in which $A_{\sigma \eta} (\e,t)$ is being evaluated. 
The others quantities are defined as
\begin{equation}
\a_\eta=V^0_\eta-V^0_d,
\end{equation}
\begin{equation}
 \g(t,t_1)=e^{i( \e + i \frac{\G_\s}{2} -\e_0)(t-t_1) },
\end{equation}
and $\xi=0$ for $N=0$ and $\xi=1$ for $N \ge 1$.
It is yet valid to mention that $\D_\eta$ is assumed constant throughout the leads. Physically, this means that the electronic system has enough time to screen an external electric field as it evolves in time. This assumption implies bias modulations not faster than the typical plasma frequency.\cite{nsw93} The three integrals in Eq. (\ref{Awithint}) can be solved analytically, thus resulting in the following expression
\begin{eqnarray}\label{Afinal}
&&A_{\s \eta}(\e,t)=\frac{e^{i(\e-\e_0+i\frac{\G_\s}{2})t} e^{-i \a_\eta [t \phi(t) + \sum_{n=0}^{N} (-1)^{n+1}s_n]}}{\e-\e_0+i \frac{\G_\s}{2}}+\nonumber\\
&&\phantom{xxx}\xi \sum_{n=0}^{N-1} e^{i(\e-\e_0+i \frac{\G_\s}{2})t} e^{-i \a_\eta [t \phi(t) + \sum_{l=n+1}^{N} (-1)^{l+1} s_l]} \times \nonumber \\
&&\phantom{xxx}\frac{[e^{-i(\e-\e_0+i \frac{\G_\s}{2}-\a_\eta f_n)s_{n+1}}-e^{-i(\e+i \frac{\G_\s}{2}-\a_\eta f_n) s_n}]}{\e-\e_0+i \frac{\G_\s}{2}-\a_\eta f_n}+\nonumber\\
&&\phantom{xxxxxxxxxxx}\frac{1-e^{i[\e-\e_0+i \frac{\G_\s}{2}-\a_\eta \phi(t)](t-s_{N})}}{\e-\e_0+i \frac{\G_\s}{2}-\a_\eta \phi(t)}.
\end{eqnarray}
In this last equation $f_n=1$ for $n$ even and $0$ for $n$ odd. Using Eq. (\ref{Afinal}) into Eqs. (\ref{Iseta}) and (\ref{nst}) we determine the dynamics of the spin polarized current and the dot occupation, as described in Sec. 4.

\section{Parameters}

In our numerical calculations we have described the ferromagnetic leads via spin dependent tunneling rates $\G_\s^\eta$, given by\cite{wr01}
\begin{eqnarray}
 \G_\s^L&=&\G_0[1+(-1)^{\d_{\s \downarrow}}p_L],\\
 \G_\s^R&=&\G_0[1\pm(-1)^{\d_{\s \downarrow}}p_R],
\end{eqnarray}
where $\G_0$ is the lead-dot coupling strength, $p_L$ and $p_R$ are the left and right lead polarization, and the $+$ $(-)$ sign in $\G_\s^R$ corresponds
to parallel (antiparallel) magnetic alignment of the leads. In particular, in the present work we assume $p_L=p_R=0.4$. The others quantities involved
are the dot's level $\e_0=5 \G_0$, the voltages intensities $V_L^0=0$, $V_R^0=20 \G_0$ and $V_d^0=10 \G_0$, the pulse width and the inverval between pulses, $s_n-s_{n-1}=3 \hbar/\G_0$, and the temperature $k_B T= \G_0$. In what follows we present the results.

\section{Results}

\begin{figure}[h]
\par
\begin{center}
\epsfig{file=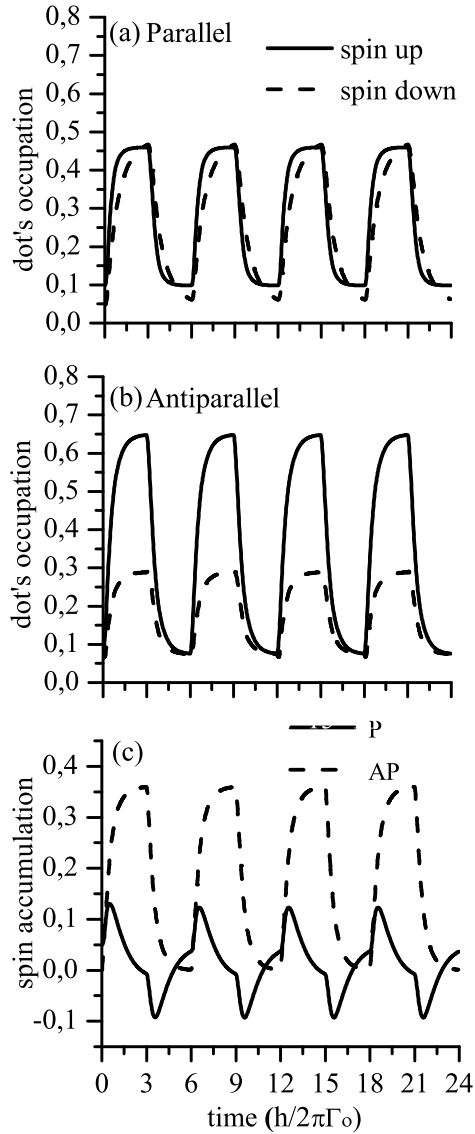, width=0.4\textwidth}
\end{center}
\caption{Dynamical spin-resolved electronic populations in the (a) parallel and (b) antiparallel configurations and (c) the spin accumulation $m=n_\uparrow - n_\downarrow$. While in the parallel configuration both $n_\uparrow$ and $n_\downarrow$ attain equal plateaus in the stationary limit, in the antiparallel case we find $n_\uparrow > n_\downarrow$ in this limit, which results in a higher amplitude of the spin accumulation wave in the AP than in the P alignment. Additionally, the spin accumulation wave changes sign in the P case.} \label{fig2}
\end{figure}

\begin{figure}[h]
\par
\begin{center}
\epsfig{file=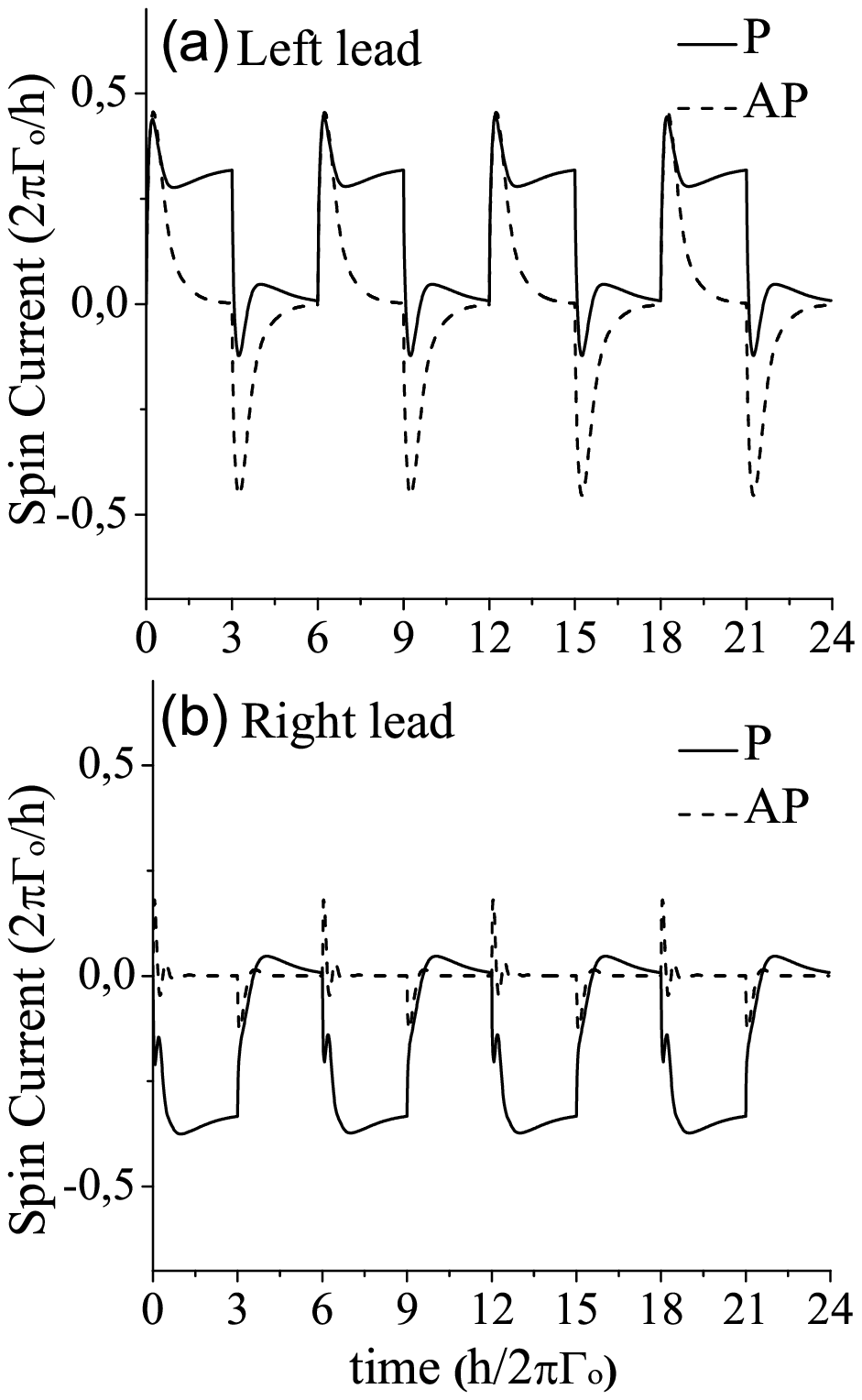, width=0.4\textwidth}
\end{center}
\caption{Spin current agains time in both parallel and antiparallel alignments and in both ferromagnetic leads. In the left lead (emitter) $I_{spin}^P$ develops higher values than $I_{spin}^{AP}$ inside the bias pulse. In contrast, when the bias voltage is turned off $I_{spin}^{AP}$ attains much higher negative values than $I_{spin}^P$.}\label{fig3}
\end{figure}

\subsection{Spin accumulation}

Figures \ref{fig2}(a)-(b) show the spin resolved dot occupations in both (a) P and (b) AP alignments. When the bias is turned on ($t=0,6\hbar/\G_0,12\hbar/\G_0,...$) the occupations increase in time due to the resonant condition $\mu_R < \e_d(t) < \mu_L(t)$ that is achieved inside the pulse length. In the parallel configuration $n_\uparrow$ presents a stepper enhancement compared to $n_\downarrow$. This is related to the inequality $\G_\uparrow^L > \G_\downarrow^L$, that gives a faster response for the incoming up spins. As the time evolves $n_\uparrow$ and $n_\downarrow$ tend to the same value, inside the pulse. In contrast, in the AP alignment $n_\uparrow$ tends to saturate above $n_\downarrow$. These asymptotic behaviors can be easily understood in terms of the tunneling rates in both alignments. While in the P case $\G_\s^L=\G_\s^R$, thus resulting in $n_\uparrow = n_\downarrow$ in the stationary limit, in the AP configuration $\G_\uparrow^L > \G_\uparrow^R$ and $\G_\downarrow^L < \G_\downarrow^R$, which gives rise to $n_\uparrow > n_\downarrow$ in the stationary limit. When the bias voltage is turned off ($t=3\hbar/\G_0,9\hbar/\G_0,15\hbar/\G_0,...$) the level $\e_d$ becomes above $\mu_L$ and $\mu_R$, thus the spin populations in the dot discharge toward the leads. The minimum values achieved by $n_\uparrow$ and $n_\downarrow$ during the discharge process come from thermal excitation. Note that just before a pulse of bias voltage we find $n_\uparrow > n_\downarrow$ in the P case while $n_\uparrow = n_\downarrow$ in the AP configuration. This is related to the broadening of the dot level $\e_d$ that is spin dependent, even though it is spin degenerate. Since in the P case $\G_\uparrow^L+\G_\uparrow^R > \G_\downarrow^L + \G_\downarrow^R$ the level broadening for spin up is larger than for spin down. This makes thermal excitations more pronounced for spins up than for spins down, thus resulting in an equilibrium spin accumulation. In the AP alignment since $\G_\uparrow^L+\G_\uparrow^R = \G_\downarrow^L + \G_\downarrow^R$ the thermal excitation is equally distributed for both spins. In Fig. \ref{fig2}(c) we show the spin accumulation in the dot, $m=n_\uparrow-n_\downarrow$. For AP configuration $m$ changes intensity preserving its sign. In contrast, in the P alignment $m$ oscillates between positive and negative values as the time evolves. The sign reversion of $m$ in the P case comes from the faster charge/discharge of spin up electrons in the dot, compared to the spin down electrons. 

\subsection{Spin current}

Figure \ref{fig3}(a) shows the spin currents ($I_{spin}=I_\uparrow-I_\downarrow$)\cite{noteSpincurr} in the emitter (left) and collector (right) lead in both magnetic alignments. Observe that just after a bias voltage is turned on, both $I_{spin}^P$ and $I_{spin}^{AP}$ are on top of each other. As the time evolves $I_{spin}^{AP}$ is suppressed to zero while $I_{spin}^{P}$ tends to a nonzero stationary value. In contrast, when the bias voltage is turned off, $I_{spin}^{AP}$ assumes transient values higher in modulos than $I_{spin}^{P}$. The negative values of the spin currents just after a pulse end means that the spin current is discharging into the emitter lead with a majority up component.\cite{foot1} In contrast, in the collector lead [Fig. \ref{fig3}(b)]  $I_{spin}^{AP}$ is aproximatelly zero throughout time, with small oscillations (ringing like) whenever a bias voltage is turned on or off. In contrast, $I_{spin}^P$ assumes relatively high negative values. Comparing Figs. \ref{fig3}(a) and \ref{fig3}(b) we note that in a transient regime the spin current that leaves the emitter does not arrive in the collector lead. Similarly to the total current ($I=I_\uparrow+I_\downarrow$) that has a continuity equation given by
\begin{equation}
 I^L+I^R=e \frac{d n}{dt},
\end{equation}
where $n=n_\uparrow+n_\downarrow$, the total spin current satisfies the following continuity equation 
\begin{equation}\label{spincontinuity}
 I_{spin}^L+I_{spin}^R = e \frac{d m}{dt},
\end{equation}
which was used to check the accuracy of our numerical results.

\begin{figure}[tbp]
\par
\begin{center}
\epsfig{file=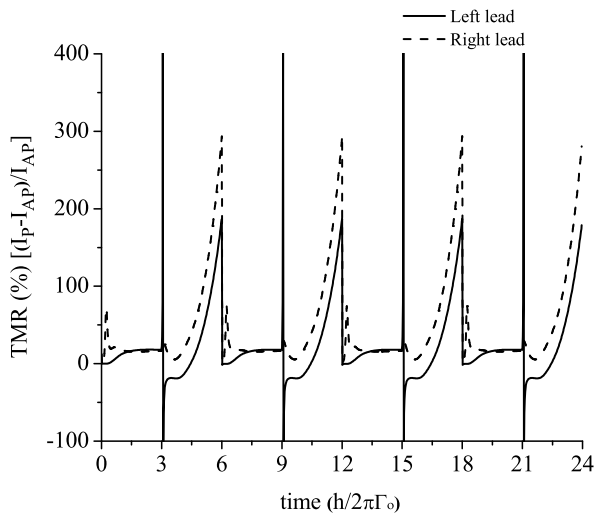, width=0.5\textwidth}
\end{center}
\caption{Tunnel magnetoresistance against time in the left (solid line) and right 
(dotted line) leads. In the time range in which $V(t)=0$ the TMR increases
up to 200-300\% due to the relatively slow discharging of spin down electrons in the dot in the
parallel alignment. This sustains $I^P$ much longer than $I^{AP}$, thus enhancing
the TMR in time. Additionally, when the bias voltage is turned off, a singularity in
the TMR is observed. This is related to the zero value attained by the current in the left
lead when it is passing from forward (positive) to reverse (negative) current.}\label{fig4}
\end{figure}

\subsection{TMR}

Figure \ref{fig4} shows the TMR$=(I^P-I^{AP})/I^{AP}$ against time in the left (solid line) and right (dotted line) leads. Here $I^P$ and $I^{AP}$ means the total current ($I_\uparrow+I_\downarrow$) in the parallel and antiparallel configurations, respectively. In the time range in which $V(t)=0$ the TMR increases up to 200-300\% due to the relatively slow discharging of spin down electrons from the dot into the leads in the parallel configuration. This sustains $I^P$ much longer than $I^{AP}$, thus enhancing
the TMR along the time. Note that in the parallel case the spin down electron in the dot is weakly coupled to both leads (majority spin up population in both sides), while in the antiparallel case both spin components are strongly coupled to at least one electrode. This turns the spin down discharging process slower in the parallel case, thus resulting in a maintenance of the total current $I^P$ for longer times compared to $I^{AP}$. It is valid to mention that whenever a bias voltage is turned off the TMR develops a singularity as described in the next figure.

Figure \ref{fig5} shows the TMR and the total currents $I^P$ and $I^{AP}$ in a time range around $3\hbar/\Gamma_0$. The divergence and sign change of the TMR observed in this figure is related to the spin-dependent discharging process of the dot. Note that after the bias voltage is turned off (at $3\hbar/\Gamma_0$) the currents $I^P$ and $I^{AP}$
transiently pass from direct (positive) to reverse (negative) currents. During this transition the current $I^{AP}$ crosses the zero value before $I^P$ and attains a slightly more negative value than $I^P$. This turns into a singularity in the TMR and a change of its sign.

\begin{figure}[tbp]
\par
\begin{center}
\epsfig{file=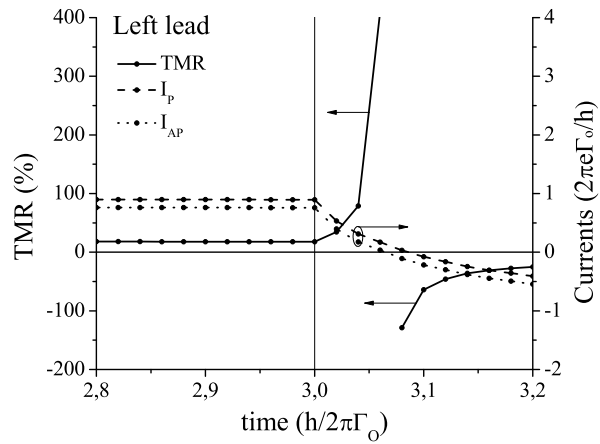, width=0.5\textwidth}
\end{center}
\caption{TMR and total currents $I^P$ and $I^{AP}$ against time around $ t = 3 \hbar / \Gamma_0 $. The TMR changes sign when $I^{AP}$ switches from forward (positive) to reverse (negative) current.} \label{fig5}
\end{figure}

\section{Conclusion}

We calculate spin polarized current, spin accumulation and TMR in a quantum dot coupled to ferromagnetic leads in the presence of a bias voltage that evolves in time as a square-wave. We report a wave like spin accumulation that switches sign in each period when the leads are parallel aligned. We also found a much larger spin current in the antiparallel configuration during the transient discharging process, which quite contrasts with the stationary regime where we find an antiparallel spin current equal zero. Finally, we find a large enhancement of the TMR in the time range in between bias voltage pulses. This is due to the slow discharge of the spin down electrons in the parallel configuration. A sign reversion of the TMR is also observed in the emitter lead, which is related to a sign change of the current.

The authors acknowledge I. Larkin, R. Zelenovsky, and A. P. Jauho for valuable comments. This work was supported by the Brazilian Ministry of Science and Technology and IBEM (Brazil).


\begin{thebibliography}{99}
\bibitem{commentUzero} It was recently pointed out that in the sequential tunneling regime
both the cases with and without Coulomb interaction give equal results when the dot can be doubly occupied, see Ref. \cite{fms07_2}.
Motivated by this, we assume that the bias voltage is sufficiently high in order to allow double
occupancy during the pulses.
\bibitem{foot1} In the text we assume that the current is positive when it leaves the leads and negative when it arrives into the leads.
\end{thebibliography}
\end{document}